\documentclass[aps,reprint,superscriptaddress,nofootinbib,longbibliography,floatfix]{revtex4-2}

\usepackage{paralist}
\usepackage{graphicx}
\usepackage{tabularx}
\usepackage{amsmath}
\usepackage{amstext}
\usepackage{amssymb}
\usepackage{xfrac}
\usepackage{relsize}
\usepackage[colorlinks,citecolor=blue]{hyperref}
\usepackage{amstext,wrapfig}
\usepackage{amsfonts}
\usepackage{longtable,booktabs}
\usepackage{url}
\usepackage{subfigure,graphicx}
\usepackage{dsfont}
\usepackage{booktabs}
\usepackage{amsbsy}
\usepackage{dcolumn}
\usepackage[inline]{enumitem}
\usepackage{amsthm}
\usepackage{bm}
\usepackage{esint}
\usepackage{multirow}
\usepackage{hyperref}
\usepackage{xurl}
\usepackage{cleveref}
\usepackage{mathrsfs}
\usepackage{amsfonts}
\usepackage{amsbsy}
\usepackage{dcolumn,physics,floatflt}
\usepackage{bm}
\usepackage{multirow}
\usepackage{color}
\usepackage{extarrows}
\usepackage{orcidlink}
\usepackage{datetime}
\usepackage{comment}
\usepackage[super]{nth}
\usepackage{tikz-cd}
\usepackage{braket}
\setlist[enumerate]{left=0pt, labelwidth=!,parsep=0pt,topsep=0pt,itemsep=0pt}

\renewcommand{\Re}{\operatorname{Re}}
\renewcommand{\Im}{\operatorname{Im}}

\begin{document}

\title{Nonunitary Quantum Criticality Driven by a General Boundary Condition: Exact Solution}

\author{Yao Zhou}
\affiliation{School of Physics, State Key Laboratory of Optoelectronic Materials and Technologies, and Guangdong Provincial Key Laboratory of Magnetoelectric Physics and Devices, Sun Yat-sen University, Guangzhou, China}
\affiliation{Department of Physics and HK Institute of Quantum Science \& Technology, The University of Hong Kong, Hong Kong, China}

\author{Peng Ye}
\email{yepeng5@mail.sysu.edu.cn}
\affiliation{School of Physics, State Key Laboratory of Optoelectronic Materials and Technologies, and Guangdong Provincial Key Laboratory of Magnetoelectric Physics and Devices, Sun Yat-sen University, Guangzhou, China}

\date{\today}

\begin{abstract}
Nonunitary quantum critical phenomena remain largely unexplored, hindered by the scarcity of exactly solvable models and the numerical challenges in non-Hermitian systems. Here we demonstrate that a general boundary condition drives fundamentally new nonunitary boundary quantum criticality in an exactly solvable lattice model. The exact solution yields a closed-form expression for the boundary-induced equal-time correlation function. The algebraic long-distance behavior of the boundary-induced correlation is governed by the single-particle reflection amplitude at the Fermi level, whose anomalous amplification is forbidden in unitary scattering and therefore provides a direct scattering signature of the nonunitary boundary-critical regime. We use the biorthogonal entanglement entropy as a many-body probe of this boundary-critical behavior and find that it retains logarithmic scaling, while the associated effective central charge, continuously tunable by the boundary parameters, changes from real to complex. The real branch admits a unitary defect conformal field theory (CFT) description, whereas the complex branch lies beyond analytic continuation from the unitary theory. These results establish minimal boundary engineering as an exactly solvable platform for nonunitary boundary quantum criticality and provide a benchmark for broader studies of nonunitary critical phenomena.
\end{abstract}

\maketitle
\section{Introduction}
Recent years have witnessed remarkable progress in non-Hermitian physics, unveiling a host of unconventional phenomena with no Hermitian counterparts~\cite{ashidaNonHermitian2020,bergholtzExceptional2021}. Prominent examples include the non-Hermitian skin effect~\cite{MartinezNonHermitian2018,yaoEdge2018,yokomizoNonBloch2019,SebastianTopolgical2020,XiaononHermitian2020,helbigGeneralized2020,zhangCorrespondence2020,ZhangObervation2021,LiangDynamic2022,zhangUniversal2022,ZhaoTwo2025,HetenyiLocalized2025}, non-Hermitian delocalization~\cite{hatanoLocalization1996,HatanoNonHermitian1998,longhiTopological2019,KawabataNonunitary2021,LinTopological2022,chenQuantum2022,GoncalvesCritical2023,ZhangNonHermitian2024,LonghiErratic2025,JinAnderson2025}, exceptional points~\cite{heissPhysics2012,zhenSpawning2015,hahnObservation2016,zhangObservation2017,miriExceptional2019,wangArbitrary2019,xiaoObservation2021}, and nonunitary critical phenomena~\cite{ChangEntanglement2020,HaldarHidden2023,JacobsenLattice2024,TangReclaiming2024}. These discoveries have broadened our understanding of quantum matter and motivated extensions of existing theoretical frameworks.

A theoretical foundation for nonunitary quantum criticality was established by the Yang--Lee theory of the Ising model in an imaginary magnetic field~\cite{YangStatistical1952,LeeStatistical1952,FisherYang1978,CardyConformal1985}, described by nonunitary minimal models with a discrete series of negative real central charges~\cite{francesco2012conformal}. Since then, uniform non-Hermitian effects have become a standard route to nonunitary criticality~\cite{ChangEntanglement2020,HaldarHidden2023,JacobsenLattice2024,TangReclaiming2024}, encompassing the non-Hermitian Su-Schrieffer-Heeger model~\cite{ChangEntanglement2020} and non-Hermitian $Q$-state Potts model~\cite{GorbenkoWalking2018,GorbenkoWalking2018a,JacobsenLattice2024}. Across these examples, the effective central charges realized on the lattice are fixed and discrete~\cite{ChangEntanglement2020,JacobsenLattice2024,TangReclaiming2024}: most are purely real, while the $Q$-state Potts model~\cite{GorbenkoWalking2018,GorbenkoWalking2018a,JacobsenLattice2024} and the O($n>2$) model~\cite{HaldarHidden2023} provide exceptions with discrete complex values. Continuously tunable complex central charges, by contrast, have appeared only in continuum theories such as the complex Liouville string~\cite{CollierComplex2025}. Beyond the nonunitary minimal models, whose central charges follow from an exact algebraic formula, uniform non-Hermitian effects greatly impede analytical treatment: central charges are obtained almost exclusively numerically, and even in free-fermion systems an analytical result remains elusive~\cite{ChangEntanglement2020,LiImpurity2025}.

Alternatively, modifying boundary conditions or engineering defects offers a complementary route to manipulating critical behavior~\cite{andreiBoundary2020a,geDefect2024,LiImpurity2025}.
Boundary phenomena have been widely studied in topological physics~\cite{HasanColloquium2010,QiTopological2011,ArmitageWeyl2018}, quantum criticality~\cite{CardyBoundary1989,cardy1996scaling,DiehlTheory1997,andreiBoundary2020a}, and the AdS/bCFT correspondence~\cite{TakayanagiHolographic2011,FujitaAds2011}.
Yet these studies predominantly assume boundaries that preserve Hermiticity. Nonunitary critical phenomena induced by a \emph{general boundary condition} (or equivalently, a general defect) thus remain largely unexplored, partly because the validity of analytic continuation from unitary defect/boundary CFTs is highly questionable~\cite{GorbenkoWalking2018,GorbenkoWalking2018a,andreiBoundary2020a,HaldarHidden2023,JacobsenLattice2024,TangReclaiming2024}.
In such a setting, only the boundary introduces non-Hermiticity while the bulk remains fully Hermitian, \emph{in sharp contrast to} previous works on non-Hermitian bulk critical phenomena~\cite{ChangEntanglement2020,LeeExceptional2022,HaldarHidden2023,JacobsenLattice2024,TangReclaiming2024,ShimizuComplex2025}.
This raises a fundamental open question: can a \emph{general boundary condition} on a Hermitian gapless system induce nonunitary critical phenomena with universal scaling behaviors beyond the reach of unitary defect CFTs?

\begin{figure}[htbp]
\centering
\includegraphics[width=8cm]{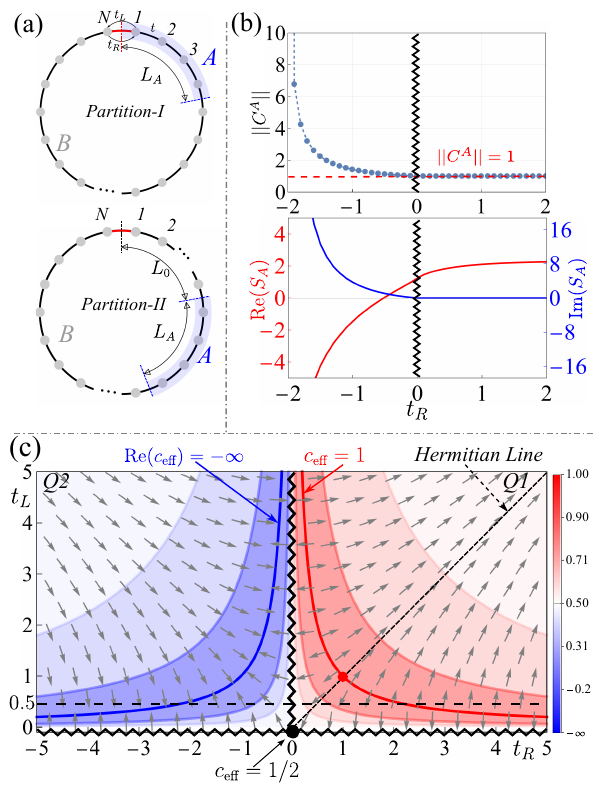}
\caption{
(a) Model Hamiltonian $\hat{H}_{d}$: the red bond marks the general boundary ($t_{R},t_{L}$), black bonds denote uniform Hermitian hopping $t$.
Partition-I and Partition-II cut as indicated; $L_{0}$ is the boundary-subsystem distance and $L_{A}$ the subsystem size.
(b) Norm of the subsystem correlation matrix $\left\lVert C^{A} \right\rVert$ and real/imaginary parts of biorthogonal EE $S_{A}$ for subsystem $A$ ($L_{A}=100$) as functions of $t_{R}$ at fixed $t_{L}=0.5$ using Partition-I.
(c) Effective central charge $c_{\text{eff}}$ versus boundary parameters $t_{R}$ and $t_{L}$ for Partition-I.
In the first quadrant (Q1: $t_{R},t_{L}>0$), $c_{\text{eff}}$ is real; in the second (Q2: $t_{R}<0$, $t_{L}>0$), it becomes complex and we plot $\mathrm{Re}(c_{\text{eff}})$.
Along $t_{R}t_{L}=t_{\text{eff}}^{2}$, $c_{\text{eff}}$ stays constant; the dashed line denotes the Hermitian limit with symmetric hopping $t_{\text{eff}}$.
}
\label{fig_model_illust}
\end{figure}

To answer this question, we exactly solve a lattice model with the general boundary condition [Fig.~\ref{fig_model_illust}(a)] and obtain the long-distance asymptotic form of the equal-time two-point correlation function:
\begin{equation}
    C(l,m) \approx \frac{\sin[\frac{\pi}{2}(l-m)]}{\pi(l-m)} -\frac{1-t_{R}t_{L}}{1+t_{R}t_{L}} \frac{\sin[\frac{\pi}{2}(l+m)]}{\pi(l+m)}.
    \label{eq_corr_asym_key}
\end{equation}
The second term contains all boundary-dependent information in the long-distance correlation function, thus encoding possible anomalous boundary-critical behavior. We next examine how the boundary affects critical behavior by analyzing the corresponding non-Hermitian single-particle scattering problem. A direct scattering calculation gives the Fermi-level reflection amplitude $r=(1-t_Rt_L)/(1+t_Rt_L)$, precisely the factor governing this term (see Appendix \ref{appendixD}). This result therefore establishes the first exact microscopic connection between boundary-induced correlations and reflection for a non-Hermitian boundary of a Hermitian gapless bulk, a setting distinct from previous field-theoretical treatments of reflection at conformal boundaries and interfaces~\cite{CardyBoundary1989,Oshikawa1996dj,quellaReflection2007}.

Based on this connection, we characterize the boundary-induced critical behavior across different parameter regimes using the biorthogonal reflection and transmission coefficients. These coefficients obey the conservation relation $\mathcal R_{\mathrm{bi}}+\mathcal T_{\mathrm{bi}}=1$~\cite{XuPseudoHermiticity2023,RyuBiorthogonal2026}. For the real boundary parameters in our paper, the biorthogonal reflection coefficient is $\mathcal R_{\mathrm{bi}}=r^{*}r=r^{2}$.
In Q1 of Fig.~\ref{fig_model_illust}(c), $0\leq\mathcal R_{\mathrm{bi}}<1$ and $0<\mathcal T_{\mathrm{bi}}\leq1$, so the scattering behavior resembles conventional unitary scattering. Consistently, the equal-time correlation function is identical to that of a Hermitian boundary with the real effective hopping $t_{\mathrm{eff}}=\sqrt{t_Rt_L}$, further indicating that the boundary-induced critical behavior in Q1 is unitary-like.
In Q2, no corresponding Hermitian mapping exists, while $\mathcal R_{\mathrm{bi}}>1$ and $\mathcal T_{\mathrm{bi}}<0$, beyond the bounds of unitary scattering. These coefficients provide a direct microscopic scattering criterion for the nonunitary boundary-critical regime, with $\lVert C^{A}\rVert>1$ [Fig.~\ref{fig_model_illust}(b)]. At $t_Rt_L=-1$, both $r$ and $\mathcal R_{\mathrm{bi}}$ diverge, corresponding to a singular reflection resonance at the Fermi level.

We then use biorthogonal entanglement entropy (EE) as a many-body scaling probe to extract critical information in these two parameter regimes. This scaling approach has been widely used for this purpose in various non-Hermitian systems~\cite{ChangEntanglement2020,GuoEntanglement2021,liuNonHermitian2024,ChenQuantum2024,LiImpurity2025,ShimizuComplex2025}. Across the two regimes, the biorthogonal EE retains logarithmic scaling, while $c_{\text{eff}}$ changes from real in Q1 to complex in Q2. We obtain an exact expression for $c_{\text{eff}}$ in Q1 and a concise analytical formula that accurately captures the numerical results in Q2. Unlike the discrete, fixed central charges found in representative nonunitary lattice CFTs~\cite{ChangEntanglement2020,HaldarHidden2023,JacobsenLattice2024,TangReclaiming2024}, the effective central charge obtained here is continuously tunable via the boundary parameters. 

The paper is organized as follows. In Sec.~\ref{section2}, we present the lattice model and its exact correlations and boundary scattering.  In Sec.~\ref{section3}, we study the analytic continuation from unitary defect CFTs.  In Sec.~\ref{section4}, we study the nonunitary boundary-critical regime. A short discussion is given in Sec.~\ref{section5}. The attached appendices include technical details:  the analytical solution of
  the model $\hat{H}_{d}$ (Appendix \ref{appendixA});  The bi-normalization of the right and left
  eigenvectors (Appendix \ref{appendixB});  The analytical expression of the correlation function (Appendix \ref{appendixC});
 The scattering interpretation of the boundary-induced correlation and
  biorthogonal conservation (Appendix \ref{appendixD});  The bound states of a $1$D chain with a
  general boundary condition (Appendix \ref{appendixE}).

\section{Lattice model, exact correlations, and boundary scattering}\label{section2}
We consider a free-fermion chain with a Hermitian bulk and a single general boundary condition: $\hat{H}_{d}= \sum_{n,m} \hat{c}^{\dagger}_{n} \mathcal{H}_{n,m} \hat{c}_{m} = \hat{H}_{0} + \hat{h}_{d}$, where the Hermitian part is given by $\hat{H}_{0}= -t \sum_{n=1}^{N-1} (\hat{c}^{\dagger}_{n} \hat{c}_{n+1} + \hat{c}^{\dagger}_{n+1} \hat{c}_{n}) - \mu \sum_{n=1}^{N} \hat{c}^{\dagger}_{n} \hat{c}_{n}$, and the boundary term connecting sites 1 and $N$ by $\hat{h}_{d} = -t_{R} \hat{c}^{\dagger}_{1} \hat{c}_{N} - t_{L} \hat{c}^{\dagger}_{N} \hat{c}_{1}$. Here, $\hat{c}^{\dagger}_{n}$ ($\hat{c}_{n}$) is the fermionic creation (annihilation) operator, $t$, $t_{R}$, $t_{L} \in \mathbb{R}$ are hopping amplitudes, $\mu$ is the chemical potential, and $N$ is the lattice size. We set the bulk hopping $t=1$ as the energy unit. When $t_{R} = t_{L}$, the boundary term becomes Hermitian, as illustrated by the dashed black line in Fig.~\ref{fig_model_illust}(c).
The limits $t_{R} = t_{L} = 0$ and $t = t_{R} = t_{L}$ respectively correspond to open and periodic boundary conditions. We solve the model exactly in Appendices \ref{appendixA} and \ref{appendixB}, obtaining its eigenvalues and left and right eigenstates.

We next use the equal-time correlation function to investigate the boundary-induced critical behavior, since its long-distance asymptotic behavior reflects the features of the critical behavior. For non-Hermitian free fermions, the equal-time correlation function also directly determines the biorthogonal EE, allowing us to extract quantitative boundary critical information from biorthogonal-EE scaling~\cite{HerviouEntanglement2019,ChangEntanglement2020,ChenEntanglement2021,ChenQuantum2024}. The equal-time correlation function is defined as $C(l,m)=\langle G^{L}|\hat{c}^{\dagger}_{l}\hat{c}_{m}|G^{R}\rangle$ in non-Hermitian systems, where $|G^{R}\rangle$ and $|G^{L}\rangle$ denote the right and left many-body ground states, respectively. Within the biorthogonal formulation, both left and right eigenstates are required for a complete description of a non-Hermitian system~\cite{ashidaNonHermitian2020,bergholtzExceptional2021}. Using the exact eigenvectors obtained in the appendix, we derive $C(l,m)$ in the thermodynamic limit as
\begin{equation}
C(l,m) = \frac{N}{2\pi} \int_{-k_{F}}^{k_{F}} (\tilde{\psi}^{L}_{k}(l))^{*} \tilde{\psi}^{R}_{k}(m) \, dk,
\end{equation}
where $k$ labels the eigenvectors $\tilde{\psi}^{R,L}_{k}$ with $z = e^{ik}$, $k_{F}$ is the Fermi wave vector,  and $k$ is shifted from the interval $[0,2\pi)$ to $[-\pi,\pi)$ without loss of generality. This integral can be further simplified to (see Appendix~\ref{appendixC})
\begin{equation}
C(l,m) = \frac{\sin[k_{F}(l - m)]}{\pi(l - m)} + C_{d}(l,m).
\label{eq_correla_matx}
\end{equation}
where $k_{F} = \arccos(-\mu/2) = \pi/2$ for $\mu = 0$, and states with $E_{k} < 0$ are filled. The first term in Eq.~\eqref{eq_correla_matx} is the sine kernel, which preserves translational invariance and satisfies $\lim_{l\rightarrow m}\frac{\sin[\frac{\pi}{2}(l - m)]}{\pi(l - m)} = 1/2$. The asymptotic behavior of its determinant plays a key role in analyzing the entanglement scaling in various systems~\cite{peschelEntanglement2005,SusstrunkFree2012,eislerFreefermion2013,ossipovEntanglement2014}.
The second term, $C_{d}(l,m)$, is the boundary-induced correlation that breaks translational invariance:
\begin{equation}
    C_{d}(l,m) =
    \frac{(1-t_{L}t_{R})}{4\pi}\int_{-\frac{\pi}{2}}^{\frac{\pi}{2}} 2 \Re(F(k))d k,
    \label{eq_correlad}
\end{equation}
where $\Re(F(k))=\Re[e^{-i (l+m)k}/(t_{L}t_{R}e^{- 2 i k}-1)]$. If $t_{L}t_{R}$ is real, then $C_{d}(l,m)$ is also real and satisfies $C_{d}(l,m) = [C_{d}(m,l)]^{*}$. Consequently, restricting the thermodynamic-limit correlation kernel $C(l,m)$ to any finite subsystem yields a Hermitian correlation matrix.

To illustrate the significance of the analytical expressions in Eqs.~\eqref{eq_correla_matx} and~\eqref{eq_correlad}, we consider some specific limits. First, for $t_{L}t_{R}=1$, $C_{d}(l,m)=0$, so $C(l,m)$ consists solely of the sine kernel. Consequently, the biorthogonal EE scales as $S_{A}\sim\frac{1}{3}\log L_{A}$ under both Partition-I and Partition-II. At $t_{L}=t_{R}=1$, this result reproduces the Hermitian case represented by the red point in Fig.~\ref{fig_model_illust}(c) and studied in Refs.~\cite{ossipovEntanglement2014,pouranvariEffect2015}.
Second, for $t_{L} t_{R} = 0$, Eq.~\eqref{eq_correlad} simplifies to $C_{d}(l,m) = -\frac{\sin\left[\frac{\pi}{2}(l + m)\right]}{\pi(l + m)}$. Under this condition, the biorthogonal EE scales as $S^{I}_{A} \sim \frac{1}{6} \log L_{A}$ under Partition-I, as indicated by the black point and wavy line in Fig.~\ref{fig_model_illust}(c). At $t_{L} = t_{R} = 0$, this result reproduces the Hermitian open-boundary case represented by the black point in Fig.~\ref{fig_model_illust}(c) and studied in Refs.~\cite{calabreseEntanglement2012,PeschelExact2012,eislerFreefermion2013}.

For $t_{R} t_{L} \neq 0,1$, we evaluate $C_{d}(l,m)$ exactly using contour integral techniques (see  Appendix~\ref{appendixC}). The result is $C_{d}(l,m) = b(l,m)$ for $l = m$, and $C_{d}(l,m) = f(t_{R}, t_{L})\, a(l,m) + b(l,m)$ for $l \neq m$, where $f(t_{R}, t_{L}) = {}_{2}F_{1}\left(1, \frac{l+m}{2}; \frac{l+m}{2}+1; -t_{R} t_{L}\right)$ is the hypergeometric function~\cite{NISTDLMF}, $a(l,m) = \frac{ (t_{R} t_{L}-1)\sin[\frac{\pi}{2}(l+m)]}{\pi (l+m)}$, and $b(l,m) = \frac{1 - t_{R} t_{L}}{2} (t_{R} t_{L})^{-\frac{l+m}{2}}$ for $|t_{R} t_{L}| > 1$ (while $b(l,m) = 0$ for $|t_{R} t_{L}| < 1$).

For $l + m \gg 1$, the hypergeometric function has the asymptotic form $f(t_{R}, t_{L}) \approx \frac{1}{1 + t_{R} t_{L}}$~\cite{NISTDLMF}. Consequently, Eq.~\eqref{eq_correla_matx} simplifies to
\begin{equation}
    C(l,m) \approx \frac{\sin[\frac{\pi}{2}(l-m)]}{\pi(l-m)} -r \frac{\sin[\frac{\pi}{2}(l+m)]}{\pi(l+m)} + b(l,m).
    \label{eq_corr_asym}
\end{equation}
where $r=(1-t_{R}t_{L})/(1+t_{R}t_{L})$.

To understand how the boundary influences critical behavior, we analyze the corresponding non-Hermitian single-particle scattering problem in Appendix~\ref{appendixD}. The calculation identifies the coefficient $r$ of the boundary-induced term in Eq.~\eqref{eq_corr_asym} as the Fermi-level reflection amplitude. Based on this connection, we characterize the boundary-induced critical behavior using the biorthogonal reflection and transmission coefficients, which satisfy $\mathcal R_{\mathrm{bi}}+\mathcal T_{\mathrm{bi}}=1$~\cite{XuPseudoHermiticity2023,RyuBiorthogonal2026}. For the real boundary parameters in our paper, the biorthogonal reflection coefficient is $\mathcal R_{\mathrm{bi}}=r^{*}r=r^{2}$.

In the Hermitian limit, the biorthogonal coefficients reduce to the conventional reflection and transmission probabilities, denoted by $R=|r|^{2}$ and $T$, respectively. The scattering matrix is unitary, so $R+T=1$. Periodic boundary conditions ($t_{R}=t_{L}=1$) give $r=0$ and $R=0$, corresponding to perfect transmission, whereas open boundary conditions ($t_{R}=t_{L}=0$) give $r=1$ and $R=1$, corresponding to total reflection. Other Hermitian boundaries in the parameter range considered here ($t_{R}=t_{L}>0$, $t_{R}\neq1$) give $0<R<1$, corresponding to partial reflection.

We now turn to non-Hermitian boundaries ($t_{R}\neq t_{L}$), for which the biorthogonal scattering coefficients exhibit a clear dichotomy between Q1 and Q2. In Q1 ($t_{R},t_{L}>0$), the correlation function in Eq.~\eqref{eq_corr_asym} is identical to that of a Hermitian boundary with a real effective hopping $t_{\text{eff}}=\sqrt{t_{R}t_{L}}$, as illustrated by the dashed line in Fig.~\ref{fig_model_illust}(c). Correspondingly, $0\leq\mathcal R_{\mathrm{bi}},\mathcal T_{\mathrm{bi}}\leq1$, within the bounds of conventional unitary scattering. Thus, despite the boundary non-Hermiticity, the scattering behavior remains unitary-like in Q1.

In Q2 ($t_{R}<0$, $t_{L}>0$), however, no corresponding Hermitian mapping exists. Instead, $\mathcal R_{\mathrm{bi}}>1$ and $\mathcal T_{\mathrm{bi}}<0$. These values have no counterpart among the conventional reflection and transmission probabilities of unitary scattering. Since the same reflection amplitude controls the algebraically decaying boundary-induced correlation in Eq.~\eqref{eq_corr_asym}, the anomalous reflection in Q2 provides a microscopic scattering characterization of the nonunitary boundary-critical regime. At $t_{R}t_{L}=-1$, both $r$ and $\mathcal R_{\mathrm{bi}}$ diverge, identifying the singular point as a Fermi-level reflection resonance.

In one-dimensional gapless systems, biorthogonal EE provides a useful many-body probe of boundary-induced critical behavior~\cite{ChangEntanglement2020,andreiBoundary2020a,geDefect2024,ShimizuComplex2025}. For free-fermion non-Hermitian systems, the subsystem correlation matrix $C^{A}$ is obtained by restricting $C$ to sites $l,m\in A$. The biorthogonal EE is then calculated as $S_{A}=-\sum_{i}\left[\xi_{i}\log\xi_{i}+(1-\xi_{i})\log(1-\xi_{i})\right]$, where $\xi_i$ are the eigenvalues of $C^{A}$~\cite{peschelCalculation2003,lattorreground2004,HerviouEntanglement2019,ChangEntanglement2020}. For a Hermitian boundary, the biorthogonal EE reduces to the conventional EE. We now calculate the biorthogonal EE from Eq.~\eqref{eq_corr_asym} and use its scaling to extract critical information in the two regimes.

\begin{figure}[htbp]
\centering
\includegraphics[width=8.5cm]{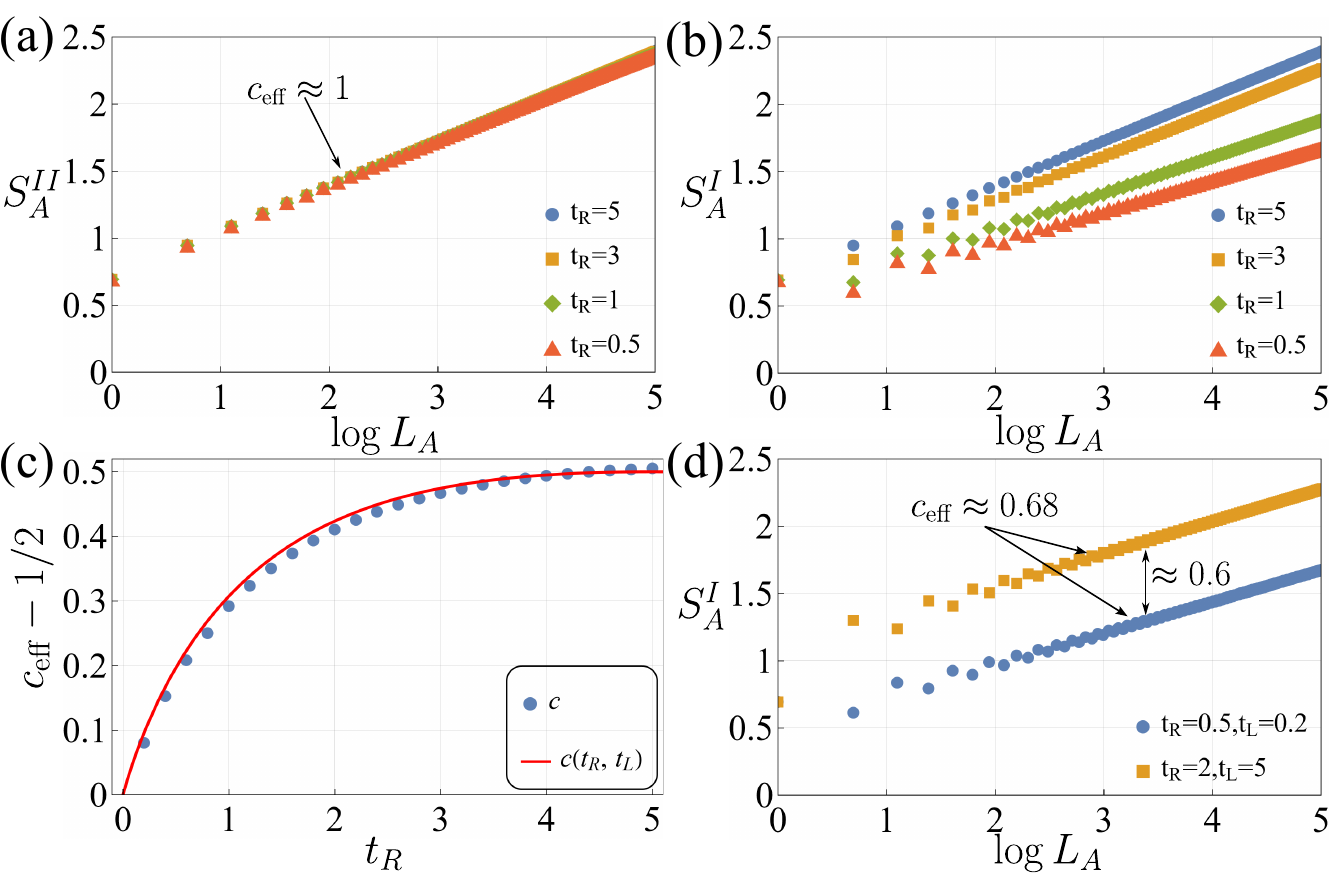}
\caption{
(a) and (b) show the logarithmic scaling of the biorthogonal EE with the varying parameter $t_{R}$ in Q1, using Partition-II and Partition-I, respectively. (c) presents $c_{\text{eff}} - 1/2$ as a function of $t_{R}$ for Partition-I,  where the blue points and red line are numerical results and fit function in Eq.~\eqref{eq_central_formu}, respectively. (d) displays the scaling of the biorthogonal EE for two sets of dual parameters, $(t_{R}, t_{L})$ and $\left( 1/t_{R}, 1/t_{L} \right)$, using Partition-I. Here, $t_{L} = 0.2$, $L_{0} = 500$ for Partition-II, and the total size of the system $N=5000$.
}\label{fig_EE_scal_first}
\end{figure}

\section{Analytic continuation from unitary defect CFTs} \label{section3}
We first examine the biorthogonal EE for $0 < t_{R}, t_{L} < 1$ in Q1 of Fig.~\ref{fig_model_illust}(c). The range $-1 < t_{R}, t_{L} < 0$ yields the same physics, as follows from Eqs.~\eqref{eq_correla_matx} and~\eqref{eq_correlad}.
For Partition-II in Fig.~\ref{fig_model_illust}(a), subsystem $A$ is separated from the boundary by a distance $L_{0}$. In this case, the second term in the asymptotic form of $C(l, m)$ in Eq.~\eqref{eq_corr_asym} satisfies $\left|\frac{\sin \frac{\pi}{2}(2L_{0} + l + m)}{\pi (2 L_{0} + l + m)}\right| < \frac{1}{2 \pi L_{0}} \rightarrow 0$ in the controlled scaling limit $1 \ll L_A \ll L_{0} \ll N$, where $N$ is the total system size. Thus, the biorthogonal EE is determined solely by the sine-kernel term of $C(l, m)$, giving $S^{II}_{A} \sim \frac{1}{3} \log L_{A}$, as in the Hermitian boundary case~\cite{ossipovEntanglement2014, pouranvariEffect2015}. This perfectly matches the numerical results in Fig.~\ref{fig_EE_scal_first}(a) and shows that the bulk remains described by a unitary CFT, despite the boundary breaking Hermiticity.

To isolate boundary effects, we consider Partition-I cutting exactly at the boundary [Fig.~\ref{fig_model_illust}(a)], and vary $t_{R}$ with $t_{L}=0.2$ fixed. The biorthogonal EE $S^{I}_{A}$ always follows $S^{I}_A = \frac{c_{\text{eff}}}{3} \log L_A + g$, with a real effective central charge $c_{\text{eff}}$, as shown in Fig.~\ref{fig_EE_scal_first}(b). From Eq.~\eqref{eq_corr_asym}, $C(l, m)$ can be obtained by analytic continuation of the Hermitian boundary case~\cite{peschelEntanglement2005}, leading to the exact result
\begin{equation}
    c_{\text{eff}}(t_{R},t_{L})= \frac{1}{2} + \frac{128}{\pi^{4}}\Theta^{2},
    \label{eq_central_formu}
\end{equation}
where $\Theta = \arctan (\sqrt{t_{R}t_{L}})\arctan(1/\sqrt{t_{R}t_{L}})$.
The numerical data (blue points) agree closely with Eq.~\eqref{eq_central_formu} (red line) in Fig.~\ref{fig_EE_scal_first}(c). The first term, $\frac{1}{2}$, comes from the normal bond interface, while the second term originates from the boundary bond in Fig.~\ref{fig_model_illust}(a). These results constitute a direct analytic continuation of the unitary defect CFT~\cite{peschelEntanglement2005}, indicating that the non-Hermiticity of the boundary is suppressed in this case.

Equation~\eqref{eq_central_formu} further shows that $c_{\text{eff}}$ is invariant under $(t_{R}, t_{L}) \leftrightarrow (1/t_{R}, 1/t_{L})$. Figure~\ref{fig_EE_scal_first}(d) confirms this duality: the biorthogonal EE for two dual parameter sets yields the same $c_{\text{eff}}$, though the constant $g$ differs by about $0.6$. For $t_{R} t_{L} > 1$, boundary-localized bound states contribute to the equal-time correlation function; these states are absent for $0<t_{R} t_{L} < 1$. The bound-state contribution is $C_{b}(l, m) = \frac{t_{R} t_{L} - 1}{2} (t_{R} t_{L})^{-\frac{l+m}{2}}$ (derived in  Appendix~\ref{appendixE}), which decays exponentially with $l+m$. Thus, $C_{b}$ modifies only the constant part of the biorthogonal EE, consistent with numerical results in Fig.~\ref{fig_EE_scal_first}(d).

\begin{figure}[htbp]
\centering
\includegraphics[width=8.5cm]{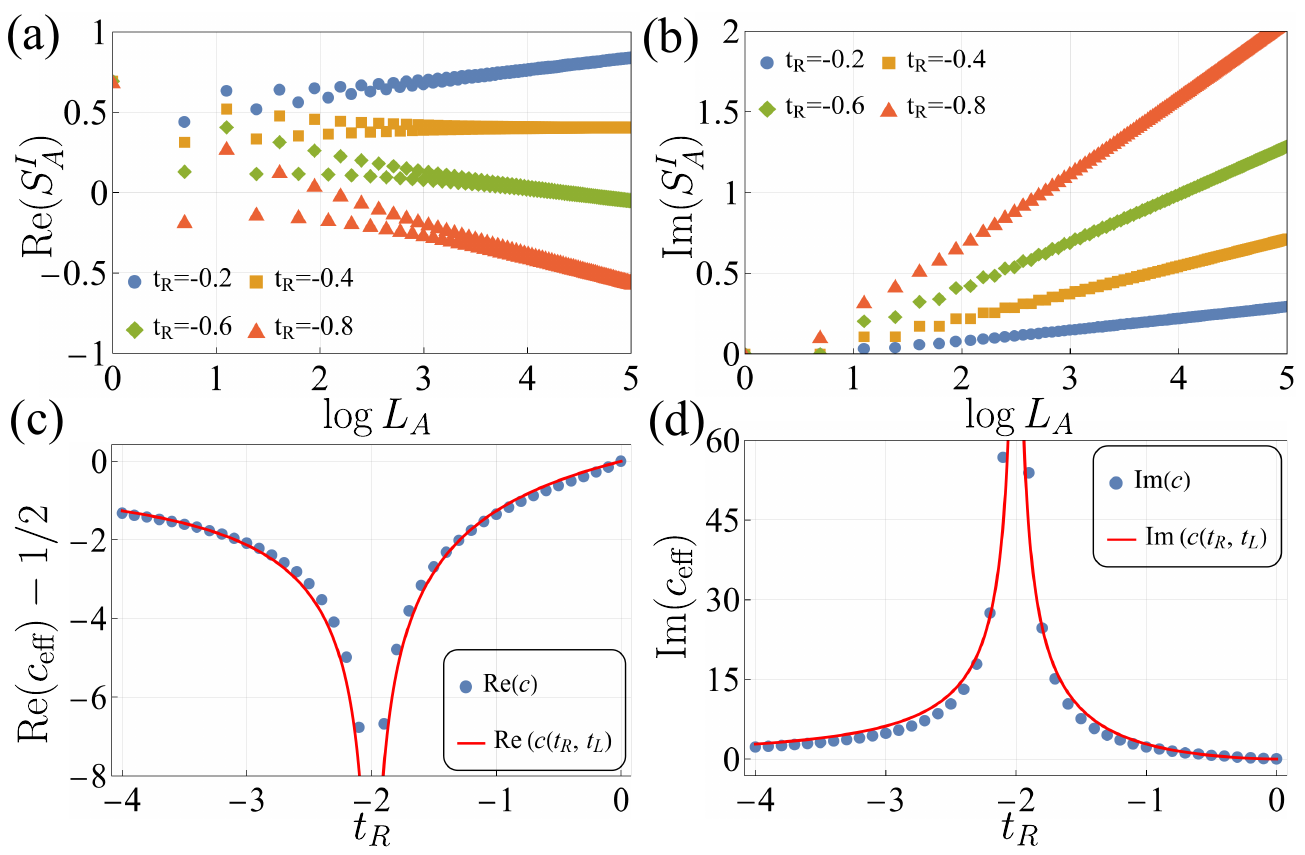}
\caption{Both the real and imaginary parts of the biorthogonal EE, $S^{I}_{A}$ in (a) and (b), exhibit logarithmic scaling with Partition-I in Q2. As $t_{R}$ decreases, the real part of $c_{\text{eff}}$ decreases, while the imaginary part increases. (c) and (d) show $\Re(c_{\text{eff}})-1/2$ and $\Im(c_{\text{eff}})$ of the phase in Q2 as functions of $t_{R}$ with $L_{A}=2048$ for Partition-I, where the blue points and red line are numerical results and fit function in Eq.~\eqref{eq_anoma_central}, respectively. Here, $t_{L} = 0.5$, and the total size of the system $N=5000$.
}\label{fig_EE_scal_second}
\end{figure}

\section{Nonunitary boundary-critical regime}\label{section4}
We now use the biorthogonal EE as a many-body probe of the Q2 regime ($t_{R}<0$, $t_{L}>0$) of Fig.~\ref{fig_model_illust}(c). Here, the factor $(1-t_{R}t_{L})/(1+t_{R}t_{L})$ in the second term of $C(l,m)$ [Eq.~\eqref{eq_corr_asym}] is strongly amplified as $t_{R}t_{L}\!\to\!-1$, reflecting the anomalous boundary response identified above. Using Partition-I, we find logarithmic scaling, $S^{I}_{A}\!\sim\!\frac{c_{\text{eff}}}{3}\log L_{A}$, with a generally complex effective central charge $c_{\text{eff}}$. The real and imaginary parts of the biorthogonal EE are shown in Figs.~\ref{fig_EE_scal_second}(a,b).

Furthermore, we numerically compute $c_{\text{eff}}$ as a function of $t_{R}$ with $t_{L}=0.5$, shown as blue points in Fig.~\ref{fig_EE_scal_second}(c) and (d). To gain analytical insight into these numerical results, one must analyze the asymptotics of the subsystem correlation matrix constructed from Eq.~\eqref{eq_corr_asym}. This matrix is a sum of a Toeplitz matrix and a Hankel matrix with independent symbols. The asymptotics of determinants of such Toeplitz$+$Hankel combinations have been solved only in special cases where the symbols satisfy specific relations~\cite{BasorEhrhardt2001,DeiftItsKrasovsky2011,GharakhlooIts2020}.
Q2 does not satisfy these conditions, so a direct analytical derivation of $c_{\text{eff}}$ from Eq.~\eqref{eq_corr_asym} is not available.

We therefore ask whether the Q1 result in Eq.~\eqref{eq_central_formu} can be analytically continued into Q2 by substituting $t_{R}\!\to\!-t_{R}$. This substitution gives
$\Re(\tilde{c}_{\text{eff}}(t_{R},t_{L})) = \frac{1}{2} - \frac{8\tilde{\Theta}^{2}}{\pi^{2}} + \frac{32\tilde{\Theta}^{2}}{\pi^{4}}\text{arctanh}^{2}\sqrt{|t_{R}t_{L}|}$ for $-1<t_{R}t_{L}<0$ [and $\Re(\tilde{c}_{\text{eff}}(t_{R},t_{L})) = \frac{1}{2} - \frac{8\tilde{\Theta}^{2}}{\pi^{2}} + \frac{32\tilde{\Theta}^{2}}{\pi^{4}}\text{arccoth}^{2}\sqrt{|t_{R}t_{L}|}$ for $t_{R}t_{L}<-1$], while $\Im(\tilde{c}_{\text{eff}}(t_{R},t_{L})) = -\frac{16}{\pi^{3}}\tilde{\Theta}^{3}$, where $\tilde{\Theta}=\log[|(1-\sqrt{\abs{t_{R}t_{L}}})/(\sqrt{\abs{t_{R}t_{L}}}+1)|]$ with $t_{R}t_{L}<0$. Comparison with the numerical data in Fig.~\ref{fig_EE_scal_second}(c) and (d) shows that the imaginary part $\Im(\tilde{c}_{\text{eff}})$ accurately captures the numerical results, whereas the real part $\Re(\tilde{c}_{\text{eff}})$ deviates significantly.

Motivated by this partial match and guided by the structure of $\Re(\tilde{c}_{\text{eff}})$, we propose the following concise formula:
\begin{equation}
\Re(c_{\text{eff}}(t_{R},t_{L})) = \frac{1}{2} - \frac{4}{\pi^{2}} \tilde{\Theta}^{2},
\qquad
\Im(c_{\text{eff}}(t_{R},t_{L})) = -\frac{16}{\pi^{3}}\tilde{\Theta}^{3},
    \label{eq_anoma_central}
\end{equation}
which retains the analytic-continuation form for $\Im(c_{\text{eff}})$ but replaces the real part with a simpler expression. As illustrated in Fig.~\ref{fig_EE_scal_second}(c) and (d), Eq.~\eqref{eq_anoma_central} accurately captures the numerical results in Q2 and provides an analytical characterization of the complex effective central charge in the nonunitary boundary-critical regime~\cite{andreiBoundary2020a}.

Finally, as a complementary characterization of the two regimes, we analyze the spectral norm $\lVert C^{A} \rVert$ under Partition-I. In Q1 [Fig.~\ref{fig_model_illust}(b)], $\lVert C^{A} \rVert = 1$, while in Q2 it exceeds $1$ and diverges at $t_{R}t_{L}=-1$. This behavior is striking: in Hermitian free-fermion systems, the full correlation matrix $C$ is a projector~\cite{hughesInversionsymmetric2011,LeePosition2014}, ensuring $\lVert C^{A} \rVert\leq1$ for any subregion. Here, $C^{A}$ remains Hermitian but with $\lVert C^{A} \rVert>1$, implying that some of its real eigenvalues $\xi_{i}$ extend beyond $[0,1]$. With the logarithms in the biorthogonal EE formula evaluated on their principal branches, such eigenvalues necessarily produce complex biorthogonal EE~\cite{HerviouEntanglement2019,ChangEntanglement2020,ChenQuantum2024}, consistent with Fig.~\ref{fig_model_illust}(b) (lower panel). Thus, in the present model, $\lVert C^{A} \rVert>1$ provides a complementary signature of the nonunitary regime and explains the complex biorthogonal EE observed in Q2.

\section{Discussion}\label{section5}
We have shown that a general boundary condition can induce a nonunitary boundary-critical regime while the gapless bulk remains Hermitian. The exact solution yields boundary-induced equal-time correlations governed by the single-particle reflection amplitude. The associated biorthogonal reflection and transmission coefficients obey the conservation relation. In Q1, both coefficients lie between zero and unity, resembling conventional unitary scattering. In Q2, the reflection coefficient exceeds unity while the transmission coefficient becomes negative. This contrast distinguishes the nonunitary Q2 regime from the unitary-like Q1 regime. We then employ biorthogonal EE as a many-body probe and find that it retains logarithmic scaling, while the continuously tunable effective central charge changes from real to complex. The real regime admits a unitary defect-CFT description, whereas the complex regime is not captured by analytic continuation from the Q1 result. Instead, we obtain a concise analytical formula that accurately captures the numerical results in Q2. These findings establish a minimal boundary modification as an exactly solvable route to nonunitary boundary quantum criticality.

Several theoretical questions remain open. A direct asymptotic derivation of Eq.~\eqref{eq_anoma_central} from the Toeplitz$+$Hankel correlation kernel would clarify the origin and domain of validity of the complex logarithmic coefficient. It would also be useful to investigate whether Wannier interpolation, which constrains entanglement-spectrum asymptotics in gapped Hermitian free fermions~\cite{LeeWannier2015}, can be adapted to the present gapless system with a non-Hermitian boundary. Extending the scattering--correlation relation to multiple non-Hermitian defects or multichannel systems could further test whether amplified reflection remains a useful diagnostic of nonunitary boundary criticality. Comparing scattering amplification with eigenstate nonorthogonality, as quantified by generalized Petermann factors used to diagnose bulk and edge exceptional points~\cite{ZouDetecting2024}, could provide a complementary perspective. Boundary zero singular modes have been linked to both Toeplitz particle--loop braiding in infinite-component $BF$ theories and non-Hermitian directional amplification~\cite{LiInfinite2026}; whether this viewpoint can elucidate the reflection resonance found here remains to be investigated.

Another direction is the interplay between boundary non-Hermiticity and lattice geometry. Studies of Hermitian free fermions have revealed entanglement fractalization on fractal lattices~\cite{ZhouFractalization2024} and area-law scaling even for gapless states with finite density of states on hyperbolic lattices~\cite{HuangHyperbolic2025}. Introducing local non-Hermitian couplings into these settings would allow one to ask whether complexification affects only the coefficients of the geometry-dependent entanglement scaling, its subleading terms, or its leading functional form. The exact one-dimensional solution presented here offers a benchmark for separating effects of boundary non-Hermiticity from those of geometry.

Experimentally, exotic critical phenomena have been realized in open systems~\cite{PengExperimental2015,BrandnerExperimental2017,FrancisMany2021,NoelMeasurement2022,KohMeasurement2023,GoogleMeasurementinduced2023,GaoExperimental2024}, such as the observation of Yang-Lee quantum criticality on a single-photon platform~\cite{GaoExperimental2024}. Such platforms provide a promising route to probe boundary-induced nonunitary criticality by locally engineering gain and loss in photonic or phononic lattices~\cite{GaoExperimental2024,linMeasuring2024}, or by introducing local measurements in quantum circuits~\cite{DiEntanglement2024}. Our results further identify anomalous reflection amplification and its singular Fermi-level resonance as direct scattering signatures of this boundary-induced nonunitary regime, with entanglement scaling serving as a complementary many-body probe.

\acknowledgments 
 We thank Y.~X.~Zhao for helpful discussions. This work was in part supported by National Natural Science Foundation of China (NSFC) under Grants No.~12474149, Research Center for Magnetoelectric Physics of Guangdong Province under Grant No.~2024B0303390001, and Guangdong Provincial Key Laboratory of Magnetoelectric Physics and Devices under Grant No.~2026B1212170049.

\appendix

 \onecolumngrid

\section{The analytical solution of the model Hamiltonian $\hat{H}_{d}$}\label{appendixA}
\subsection{The solution of the right eigen-equation}
In this part, we discuss the analytical solution of the model Hamiltonian $\hat{H}_{d}$ in detail. As discussed in the main text, we can propose an ansatz for the right wavefunction $\psi_{z}^{R}(n)=c^{R}_{1}(z^{\alpha})^{n}+c^{R}_{2}(z^{\beta})^{n}$ to satisfy two additional equations of the general boundary condition: 
\begin{equation}
  \begin{split}
-t_{R}\psi^{R}(N)-t\psi^{R}(2) &=E\psi^{R}(1), \\
-t\psi^{R}(N-1)-t_{L}\psi^{R}(1)&=E\psi^{R}(N).
  \end{split}
\end{equation}
Due to the energy dispersion $E = -t(z^{-1}+z)-\mu$, these two equations can be further simplified to:
\begin{equation}
  \begin{split}
    t \psi^{R}(0) &=t_{R}\psi^{R}(N),\\
    t_{L} \psi^{R}(1) &=t \psi^{R}(N+1).
  \end{split}
  \label{eq_def_cond}
\end{equation}
Then, inserting the ansatz right wavefunction into Eq.~\eqref{eq_def_cond}, we obtain the coefficient equation $\mathfrak{C}\cdot (c^{R}_{1},c^{R}_{2})^{T}=0$, where the coefficient matrix $\mathfrak{C}$ is written as
\begin{equation}
  \mathfrak{C}=
  \begin{pmatrix}
    t-(z^{\alpha})^{N}t_{R} & t- (z^{\beta})^{N}t_{R} \\
    -t (z^{\alpha})^{N+1}+z^{\alpha}t_{L} & -t (z^{\beta})^{N+1}+z^{\beta}t_{L}
  \end{pmatrix}.
  \label{eq_coef_mat}
\end{equation}
To always ensure the existence of a non-trivial solution for the coefficients $c_{1,2}$, we should have $\det(\mathfrak{C})=0$ concretely written as
\begin{equation}
  \begin{split}
  (z^{-1}-z)(t_{R}+t_{L})+t_{R}t_{L}(z^{-(N-1)}-z^{N-1})+(z^{N+1}-z^{-(N+1)})&=0 \\
  \frac{1}{z^{N+1}}[z^{2N}(z^{2}-t_{R}t_{L})+(1-z^{2})(t_{R}+t_{L})z^{N}+z^{2}t_{R}t_{L}-1]&=0.
  \end{split}
  \label{eq_right_constraint}
\end{equation}
Given an energy $E$, two roots of the energy dispersion satisfy $z^{\alpha}z^{\beta}=1$, then we take $z^{\alpha}=z$, and $z^{\beta}=\frac{1}{z}$ and $t=1$, where $z\neq0$.  So, we only need to solve a quadratic equation:
\begin{equation}
  a y^{2}+b y +c =0,
  \label{eq_quadra}
\end{equation}
where $y=z^{N}$, $a=z^{2}-t_{L}t_{R}$, $b=(1-z^{2})(t_{L}+t_{R})$ and $c=z^{2}t_{L}t_{R}-1$. After solving this equation, we have two self-consistent equations: 
\begin{equation}
  y_{1,2}= z^{N}.
\end{equation}
Then, we can use the solutions $z$ of the equations above to determine the energy dispersion and its associated eigenvectors. Concretely, from Eq.~\eqref{eq_coef_mat}, we have $(1-z^{N}t_{R})c^{R}_{1} + (1- z^{-N}t_{R})c^{R}_{2}=0$. So, we obtain $c^{R}_{1}=1- z^{-N}t_{R}$ and $c^{R}_{2}=-(1-z^{N}t_{R})$

\subsection{The solution of the left eigen-equation}
Since there is only a single general boundary condition in the model $\hat{H}_{d}$, we can still use the same ansatz wavefunction $\psi(n)=z^{n}$ to satisfy the equation:
\begin{equation}
  -t\psi^{L}(n)-t\psi^{L}(n+2)=E^{*}\psi^{L}(n+1), 
\end{equation}
where the lattice index $n=1,\cdots,N-2$. Then, the energy dispersion in this case is also given by $E^{*}=-t(z+z^{-1})-\mu$. 
Meanwhile,  we also use $\psi^{L}_{z}(n)=c^{L}_{1}(z^{\alpha})^{n}+c^{L}_{2}(z^{\beta})^{n}$ to satisfy two additional equations of the general boundary condition for the left eigen-equation, written as:
\begin{equation}
  \begin{split}
-t^{*}_{L}\psi^{L}(N)-t\psi^{L}(2) &=E^{*}\psi^{L}(1), \\
-t\psi^{L}(N-1)-t^{*}_{R}\psi^{L}(1)&=E^{*}\psi^{L}(N).
  \end{split}
\end{equation}
Here $z^{\alpha,\beta}$ are two roots of the energy dispersion from the left eigen-equation, still satisfying $z^{\alpha}z^{\beta}=1$. Meanwhile, like the right eigen-equation, these two equations above can also be simplified to:
\begin{equation}
  \begin{split}
    t \psi^{L}(0) &=t^{*}_{L}\psi^{L}(N),\\
    t^{*}_{R} \psi^{L}(1) &=t \psi^{L}(N+1).
  \end{split}
  \label{eq_def_cond_left}
\end{equation}
Like the discussion of the right eigenvectors above, the coefficient matrix of the left eigenvectors is written as:
\begin{equation}
  \mathfrak{C}^{L}=
  \begin{pmatrix}
    t-(z^{\alpha})^{N}t^{*}_{L} & t- (z^{\beta})^{N}t^{*}_{L} \\
    -t (z^{\alpha})^{N+1}+z^{\alpha}t^{*}_{R} & -t (z^{\beta})^{N+1}+z_{\beta}t^{*}_{R}
  \end{pmatrix}.
\end{equation}
Here we also have the constraint $\det(\mathfrak{C}^{L})=0$ explicitly written as:
\begin{equation}
  \begin{split}
  (z^{-1}-z)(t^{*}_{R}+t^{*}_{L})+t^{*}_{R}t^{*}_{L}(z^{-(N-1)}-z^{N-1})+(z^{N+1}-z^{-(N+1)})&=0 \\
  \frac{1}{z^{N+1}}[z^{2N}(z^{2}-t^{*}_{R}t^{*}_{L})+(1-z^{2})(t^{*}_{R}+t^{*}_{L})z^{N}+z^{2}t^{*}_{R}t^{*}_{L}-1]&=0.
  \end{split}
  \label{eq_left_constraint}
\end{equation}
Here we also take $z^{\alpha}=z$, and $z^{\beta}=\frac{1}{z}$ and $t=1$. 

\subsection{The relation of right and left constraint equations}
When the asymmetrical hoppings $t_{R,L}$ are real, then these two constraint conditions for right and left eigenvectors in Eqs.~\eqref{eq_right_constraint} and \eqref{eq_left_constraint} are identical. When $t_{R,L}$ are complex numbers, we need further discussion. From Eqs.~\eqref{eq_right_constraint} and \eqref{eq_left_constraint}, we have 
\begin{equation}
  P^{R}(z)=(z^{-1}-z)(t_{R}+t_{L})+t_{R}t_{L}(z^{-(N-1)}-z^{N-1})+(z^{N+1}-z^{-(N+1)})=0 
\end{equation}
and
\begin{equation}
  P^{L}(z)=(z^{-1}-z)(t^{*}_{R}+t^{*}_{L})+t^{*}_{R}t^{*}_{L}(z^{-(N-1)}-z^{N-1})+(z^{N+1}-z^{-(N+1)})=0. 
\end{equation}
We find that if $z$ is a root of the polynomial $P^{R}(z)$, then $z^{*}$ is a root of the polynomial $P^{L}(z)$, coinciding with the relation of energy spectrum $E(z)=E^{*}(z^{*})=-(z+z^{-1})-\mu$. Therefore, we can also use a similar method to obtain the coefficients of the left eigenvectors, written as $c^{L}_{1}=1- (z^{*})^{-N}t^{*}_{L}$ and $c^{L}_{2}=-(1- (z^{*})^{N}t^{*}_{L})$.
When considering the thermodynamic limit, due to $z=e^{ik}$ as discussed in the main text, the energy dispersion of $\hat{H}_{d}$ and $\hat{H}^{\dag}_{d}$ are real and identical, given as $E=-2\cos k -\mu$.

\section{The bi-normalization of the right and left eigenvectors}\label{appendixB}
By solving the coefficient matrices of the right and left eigenvectors in Eqs. ~\eqref{eq_right_constraint} and ~\eqref{eq_left_constraint}, the coefficients of eigenvectors are written as $c^{R}_{1,2}=\pm(1-z^{\mp N}t_{R})$ and $c^{L}_{1,2}=\pm(1-(z^{*})^{\mp N}t^{*}_{L})$. Then, we  obtain the right and left eigenvectors written as: 
\begin{equation}
  \begin{split}
    \psi^{R}_{z}(n)&= \frac{1}{i \sqrt{2 N}} [(1-z^{-N}t_{R})z^{n} -(1-z^{N}t_{R})z^{-n}], \\
    \psi^{L}_{z}(n)&= \frac{1}{i\sqrt{2 N}} [(1-(z^{*})^{-N}t^{*}_{L})(z^{*})^{n} -(1-(z^{*})^{N}t^{*}_{L})(z^{*})^{-n}].
  \end{split}
\end{equation}
To confirm the bi-normalization of the right and left eigenvectors, we should calculate 
\begin{equation}
  \begin{split}
  &\mathcal{N}_{LR}=\sum_{n=1}^{N}(\psi^{L}_{z}(n))^{*}\psi^{R}_{z^{'}}(n)= \\
  &\frac{1}{2N}\sum_{n=1}^{N}[(1-z^{-N} t_{L})z^{n}-(1-z^{N}t_{L})z^{-n}][(1-(z^{'})^{-N} t_{R})(z^{'})^{n}-(1-(z^{'})^{N}t_{R})(z^{'})^{-n}].
  \end{split}
\end{equation}
With further calculation, we have the bi-normalization constant $\mathcal{N}_{RL}$ represented as  
\begin{equation}
  \begin{split}
  &\mathcal{N}_{LR}= \\
  &\delta_{z,z^{'}}[(1+t_{R}t_{L}-(t_{R}+t_{L})f(N))+\frac{(t_{R}+t_{L})f(1)-f(N+1)-t_{R}t_{L}f(N-1)}{N}],
  \end{split}
\end{equation}
where the function $f(n)=\frac{(z^{-n}+z^{n})}{2}$ and $N$ is the lattice size. When considering the thermodynamic limit, i.e., $N\rightarrow \infty$, the absolute value $|z|=\sqrt[N]{|y_{1,2}|}\rightarrow 1$, then we have $f(n)=\cos n k$ and $|f(n)|\leq 1$. Then, $\mathcal{N}^{1(2)}_{LR}=\delta_{k,k^{'}}(1+t_{R}t_{L}-(t_{R}+t_{L})\cos kN)=\delta_{k,k^{'}}(1+t_{R}t_{L}-(t_{R}+t_{L})(y^{-1}_{1(2)}+y_{1(2)})/2)$, where $y_{1,2}$ are two roots of Eq.~\eqref{eq_quadra}. Then, we use $\mathcal{N}^{1(2)}_{LR}$ to bi-normalize the left and right eigenvectors as $\tilde{\psi}^{R(L)}_{z}(n) = \psi^{R(L)}_{z}(n)/\sqrt{\mathcal{N}_{LR}}$.

\section{The analytical expression of the correlation function}\label{appendixC}
Based on the definition of the equal-time correlation function $C(l,m) = \langle G^{L} | \hat{c}^{\dagger}_{l} \hat{c}_{m} | G^{R} \rangle$~\cite{peschelCalculation2003, lattorreground2004, HerviouEntanglement2019, ChangEntanglement2020}, we obtain the expression:
\begin{equation}
  C(l,m)= \frac{1}{2}\sum_{k\in\mathrm{occ}}\sum_{\lambda=1}^{2}(\tilde{\psi}^{L}_{z_{k,\lambda}}(l))^{*}\tilde{\psi}^{R}_{z_{k,\lambda}}(m),
\end{equation}
where $z_{k,1(2)}$ is the $k$th root of the equation $z^{N}=y_{1(2)}$. Meanwhile, we add $1/2$ to cancel the double-counting for the root $z_{k,\lambda}$. In the thermodynamic limit, we use $k$ and $\lambda$ instead of $z_{k,\lambda}$ to label eigenvectors. Then, we have 
\begin{equation}
  \begin{split}
  C(l,m) &= \frac{N}{4\pi}\sum_{\lambda=1}^{2}\int_{-k_{F}}^{k_{F}} (\tilde{\psi}^{L}_{k,\lambda}(l))^{*}\tilde{\psi}^{R}_{k,\lambda}(m) d k \\
  &= \frac{1}{8\pi }\int_{-k_{F}}^{k_{F}}d k\frac{1}{\mathcal{N}^{1}_{RL}}[(1-\frac{t_{L}}{y_{1}})(1-y_{1}t_{R})e^{i(l-m)k}   
  +(1-y_{1}t_{L})(1-\frac{t_{R}}{y_{1}})e^{-i(l-m)k} \\
  &-(1-\frac{t_{L}}{y_{1}})(1-\frac{t_{R}}{y_{1}})e^{i(l+m)k}
  -(1-y_{1}t_{L})(1-y_{1}t_{R})e^{-i(l+m)k}] \\
  &+\frac{1}{\mathcal{N}^{2}_{RL}}[(1-\frac{t_{L}}{y_{2}})(1-y_{2}t_{R})e^{i(l-m)k}   
  +(1-y_{2}t_{L})(1-\frac{t_{R}}{y_{2}})e^{-i(l-m)k} \\
  &-(1-\frac{t_{L}}{y_{2}})(1-\frac{t_{R}}{y_{2}})e^{i(l+m)k}
  -(1-y_{2}t_{L})(1-y_{2}t_{R})e^{-i(l+m)k}]. 
  \end{split}
\end{equation}
Further analysis decomposes the correlation function $C(l,m)$ into a translation-invariant term $C(l-m)$ and a translation-symmetry-breaking term $C(l+m)$. The former can be explicitly written as
\begin{equation}
  \begin{split}
  C(l-m) = 
  &\frac{1}{8\pi }\int_{-k_{F}}^{k_{F}}d k\frac{1}{\mathcal{N}^{1}_{RL}}[(1-\frac{t_{L}}{y_{1}})(1-y_{1}t_{R})e^{i(l-m)k}   
  +(1-y_{1}t_{L})(1-\frac{t_{R}}{y_{1}})e^{-i(l-m)k}] \\ 
  &+\frac{1}{\mathcal{N}^{2}_{RL}}[(1-\frac{t_{L}}{y_{2}})(1-y_{2}t_{R})e^{i(l-m)k}   
  +(1-y_{2}t_{L})(1-\frac{t_{R}}{y_{2}})e^{-i(l-m)k}] \\ 
  &= \frac{1}{4\pi}\int_{-k_{F}}^{k_{F}}d k [
    2 \cos (l-m)k - \\
  &\frac{2i(y_{1}y_{2}-1)(t_{L}-t_{R})[(1+y_{1}y_{2})(t_{R}+t_{L})-(1+t_{R}t_{L})(y_{1}+y_{2})]}{\mathcal{N}^{1}_{RL}\mathcal{N}^{2}_{RL}}] \\
  &= \frac{1}{2\pi}\int_{-k_{F}}^{k_{F}}d k\cos (l-m)k  = \frac{\sin(l-m)k_{F}}{\pi(l-m)}, 
  \end{split}
\end{equation}
where we use this relation $y_{1}y_{2}=\frac{c}{a}=\frac{z^{2}t_{L}t_{R}-1}{z^{2}-t_{L}t_{R}}$ and $y_{1}+y_{2}=-\frac{b}{a}=-\frac{(1-z^2)(t_L+t_R)}{z^2-t_Lt_R}$. The term $C(l+m)$ is induced by the general boundary condition, which is labeled by $C_{d}(l,m)$ in the main text and explicitly written as 
\begin{equation}
  \begin{split}
  C(l+m) = 
  &-\frac{1}{8\pi }\int_{-k_{F}}^{k_{F}}d k\frac{1}{\mathcal{N}^{1}_{RL}}[(1-\frac{t_{L}}{y_{1}})(1-\frac{t_{R}}{y_{1}})e^{i(l+m)k} +(1-y_{1}t_{L})(1-y_{1}t_{R})e^{-i(l+m)k}] \\ 
  &+\frac{1}{\mathcal{N}^{2}_{RL}}[(1-\frac{t_{L}}{y_{2}})(1-\frac{t_{R}}{y_{2}})e^{i(l+m)k} +(1-y_{2}t_{L})(1-y_{2}t_{R})e^{-i(l+m)k}] \\ 
  &= \frac{1-t_{L}t_{R}}{4\pi}\int_{-k_{F}}^{k_{F}}d k 
  \frac{e^{-i (l+m)k}}{t_{L}t_{R}e^{- 2 i k}-1} +\frac{e^{i (l+m)k}}{t_{L}t_{R}e^{2 i k}-1}. \\
  \end{split}
  \label{eq_def_indu}
\end{equation}
For generic complex $t_{L}t_{R}$, the equal-time correlation matrix is no longer Hermitian, and its entanglement spectrum and biorthogonal entanglement entropy can consequently become complex.
Also using $y_{1}y_{2}=\frac{c}{a}$ and $y_{1}+y_{2}=-\frac{b}{a}$, we can further obtain an integral expression of $C(l+m)$. This integral is powerful for analyzing the properties of the correlation function, but is difficult to evaluate in closed form.

\begin{figure}[htbp]
\centering
\includegraphics[width=8.5cm]{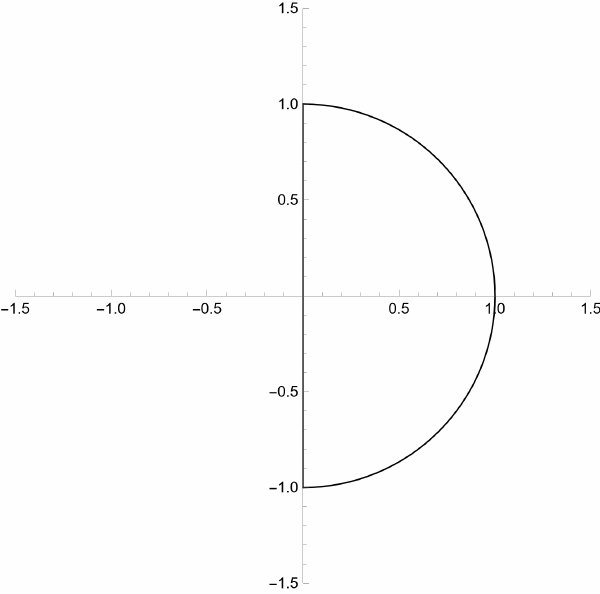}
\caption{
  The black line is the integral contour for $l+m$ being odd.
}\label{fig_integ_cont}
\end{figure}

\subsection{The analytical expression of $C(l+m)$}
In this part, we do the integral of $C(l+m)$ to obtain an analytical expression. First, we consider $k_{F}=\frac{\pi}{2}$, because of $\mu=0$. Then, we only need to consider the integral:
\begin{equation}
  F(t)=\int_{-\frac{\pi}{2}}^{\frac{\pi}{2}}\frac{e^{i(l+m)k}}{t_{R}t_{L} e^{2 i k}-1} dk,
  \label{eq_corr_part}
\end{equation}
where $C(l+m)$ becomes 
\begin{equation}
  C(l+m)=\frac{1-t_{L}t_{R}}{4\pi} 2 F(t_{R}t_{L}).
\end{equation}
When $l+m$ is even, then the integral~\eqref{eq_corr_part} can be written as a contour integral:
\begin{equation}
  2 F(t_{R}t_{L}) =\oint_{S^{1}} \frac{dz}{iz} \frac{z^{l+m}}{t_{R}t_{L}z^{2}-1}.
\end{equation}
Then, we can use the residue theorem to obtain this integral. Concretely, if $|t_{R}t_{L}|<1$, there is no singularity in the circle $S^{1}$. So, we have $2F(t_{R}t_{L})=0$. If $t_{R}t_{L}>1$, we have
\begin{equation}
  2F(t_{R}t_{L}) = 2\pi i \text{Res}[\frac{z^{l+m}}{i z(t_{R}t_{L}z^{2}-1)}] = 2\pi (t_{R}t_{L})^{-\frac{l+m}{2}}.
\end{equation}
Next, we consider $l+m$ is odd, where the integral~\eqref{eq_corr_part} can be written as: 
\begin{equation}
  \begin{split}
  F(t_{R}t_{L}) &= \oint_{\partial D} \frac{dz}{iz} \frac{z^{l+m}}{t_{R}t_{L}z^{2}-1} + \int_{-1}^{1} \frac{dy}{i y}\frac{y^{l+m} e^{i\frac{(l+m)\pi}{2}} }{t_{R}t_{L}y^{2}e^{i\pi}-1} \\
  &= \oint_{\partial D} \frac{dz}{iz} \frac{z^{l+m}}{t_{R}t_{L}z^{2}-1} - i^{l+m-1} \int_{-1}^{1} \frac{dy}{y}\frac{y^{l+m} }{t_{R}t_{L}y^{2}+1},
  \end{split}
  \label{eq_corr_odd}
\end{equation}
where $\partial D$ is the boundary line of a half circle as shown in Fig.~\ref{fig_integ_cont}.
When $|t_{R}t_{L}|<1$, the first term of Eq.~\eqref{eq_corr_odd} is $0$, due to the absence of singularity in $\partial D$. Then, we have
\begin{equation}
  F(t_{R}t_{L}) = - \sin\frac{\pi}{2}(l+m)\frac{2}{l+m} f(t_{R},t_{L}),
\end{equation}
Here $f(t_{R},t_{L})={}_{2}F_{1}\left(1, \frac{l+m}{2}; \frac{l+m}{2}+1; -t_{R} t_{L}\right)$ is the hypergeometric function \cite{NISTDLMF}. When $|t_{R}t_{L}|>1$, there is a singularity in $\partial D$. Then Eq.~\ref{eq_corr_odd} can be calculated as:
\begin{equation}
  F(t_{R}t_{L})=\pi (t_{R}t_{L})^{-\frac{l+m}{2}} - \sin\frac{\pi}{2}(l+m)\frac{2}{l+m}f(t_{R},t_{L}).
\end{equation}
Finally, based on the discussion above, when $|t_{R}t_{L}|<1$ the correlation function $C(l+m)$ can be written as 
\begin{equation}
    C(l+m) = 
    \begin{cases} 
    0     & \text{ if $l+m$ is even,} \\
    a(l,m) f(t_{R},t_{L})  & \text{ if $l+m$ is odd,} \\
    \end{cases}
    \label{eq_corr_anal1}
\end{equation}
when $|t_{R}t_{L}|>1$, $C(l+m)$ can be written as 
\begin{equation}
    C(l+m) = 
    \begin{cases} 
    b(l,m)     & \text{ if $l+m$ is even,} \\
    b(l,m)+ a(l,m) f(t_{R},t_{L}) & \text{ if $l+m$ is odd,} \\
    \end{cases}
    \label{eq_corr_anal2}
\end{equation}
where the coefficient $a(l,m)=\frac{(t_{R}t_{L}-1)\sin\frac{\pi}{2}(l+m)}{\pi(l+m)}$ and $b(l,m)=\frac{1-t_{R}t_{L}}{2}(t_{R}t_{L})^{-\frac{l+m}{2}}$. At $t_{R}t_{L}=-1$, the denominator in Eq.~\eqref{eq_corr_part} vanishes at the Fermi points, producing the singularity discussed below. At $t_{R}t_{L}=1$, although the auxiliary integral $F(t_{R}t_{L})$ is singular when considered separately, its prefactor $1-t_{R}t_{L}$ vanishes, and the complete boundary-induced correlation function is zero in the transparent-boundary limit.

\subsection{The asymptotic behavior of the hypergeometric function}
Based on Ref.\cite{NISTDLMF}, the hypergeometric function has a property:
\begin{equation}
  {}_{2}F_{1}\left(a, b; b; x\right) = (1-x)^{-a}.
\end{equation}
From the analytical expression of $C(l+m)$ in Eqs.~\eqref{eq_corr_anal1} and ~\eqref{eq_corr_anal2}, when $l+m$ is sufficiently large, we have the relation: 
\begin{equation}
{}_{2}F_{1}\left(1,\frac{l+m}{2}; \frac{l+m}{2}+1; -t_{R}t_{L}\right)
  \approx {}_{2}F_{1}\left(1,\frac{l+m}{2}; \frac{l+m}{2}; -t_{R}t_{L}\right) = (1+t_{R}t_{L})^{-1}.
\end{equation}
Therefore, $C(l+m)$ in Eqs.~\eqref{eq_corr_anal1} and ~\eqref{eq_corr_anal2} becomes as a simple expression: 
\begin{equation}
  \begin{split}
    &\text{If } |t_{R}t_{L}|<1, \\
    &C(l+m) \approx 
    \begin{cases} 
    0     & \text{ if $l+m$ is even,} \\
    -\frac{1-t_{R}t_{L}}{1+t_{R}t_{L}}\frac{\sin\frac{\pi}{2}(l+m)}{\pi(l+m)}  & \text{ if $l+m$ is odd,} \\
    \end{cases} \\
    &\text{If } |t_{R}t_{L}|>1, \\
    &C(l+m) \approx 
    \begin{cases} 
    \frac{(1-t_{R}t_{L})}{2}(t_{R}t_{L})^{-\frac{l+m}{2}}     & \text{ if $l+m$ is even,} \\
    \frac{(1-t_{R}t_{L})}{2}(t_{R}t_{L})^{-\frac{l+m}{2}}-\frac{1-t_{R}t_{L}}{1+t_{R}t_{L}}\frac{\sin\frac{\pi}{2}(l+m)}{\pi(l+m)}  & \text{ if $l+m$ is odd.} \\
    \end{cases}
  \end{split}
  \label{eq_corr_approx}
\end{equation}
Therefore, we obtain an approximate expression of the correlation function in the main text, which is highly effective for analyzing the scaling of the biorthogonal EE and testing the validity of analytic continuation from the result of a Hermitian boundary condition.

\section{Scattering interpretation of the boundary-induced correlation}\label{appendixD}
\subsection{Right scattering state}
We consider a single-particle transport setup consisting of two semi-infinite Hermitian leads connected through a generally non-Hermitian scattering region. The left and right leads are defined on sites $n\leq0$ and $n\geq1$, respectively, and have uniform hopping $t$, while the chemical potential is $\mu$ throughout. The scattering region is the asymmetric coupling between the lead endpoints $0$ and $1$, with matrix elements $\mathcal H_{1,0}=-t_R$ and $\mathcal H_{0,1}=-t_L$. We first allow $t_R$ and $t_L$ to be complex, while $t$, $\mu$, and the scattering momentum $k$ are real.

To derive the corresponding matching equations, we write a right single-particle eigenstate as
\begin{equation*}
|\psi^{\mathrm R}\rangle
=\sum_n\psi^{\mathrm R}(n)\hat c_n^\dagger|0\rangle,
\qquad
\hat H_d|\psi^{\mathrm R}\rangle=E|\psi^{\mathrm R}\rangle.
\end{equation*}
Projecting this single-particle eigenvalue equation onto sites $0$ and $1$, respectively, gives
\begin{align}
E\psi^{\mathrm R}(0)&=-t\psi^{\mathrm R}(-1)-t_L\psi^{\mathrm R}(1)-\mu\psi^{\mathrm R}(0),
\label{eq_reflection_boundary_0}\\
E\psi^{\mathrm R}(1)&=-t_R\psi^{\mathrm R}(0)-t\psi^{\mathrm R}(2)-\mu\psi^{\mathrm R}(1).
\label{eq_reflection_boundary_1}
\end{align}
For a wave incident from the left, we write
\begin{equation}
\psi^{\mathrm R}(n)=
\begin{cases}
e^{ikn}+r^{\mathrm R}(k)e^{-ikn},&n\leq 0,\\
\tau^{\mathrm R}(k)e^{ikn},&n\geq 1,
\end{cases}
\label{eq_reflection_ansatz}
\end{equation}
where $r^{\mathrm R}(k)$ and $\tau^{\mathrm R}(k)$ are the reflection and transmission amplitudes of the right scattering state, respectively. Adjacent to the scattering region,
\begin{equation}
\begin{aligned}
\psi^{\mathrm R}(-1)&=e^{-ik}+r^{\mathrm R}(k)e^{ik},\quad
\psi^{\mathrm R}(0)=1+r^{\mathrm R}(k),\\
\psi^{\mathrm R}(1)&=\tau^{\mathrm R}(k)e^{ik},\quad
\psi^{\mathrm R}(2)=\tau^{\mathrm R}(k)e^{2ik}.
\end{aligned}
\end{equation}
The scattering-state energy follows the lead dispersion $E(k)=-2t\cos k-\mu$. Substituting the expressions above into Eqs.~\eqref{eq_reflection_boundary_0} and~\eqref{eq_reflection_boundary_1} then gives
\begin{equation}
\begin{aligned}
(-2t\cos k-\mu)[1+r^{\mathrm R}(k)]
={}&-t[e^{-ik}+r^{\mathrm R}(k)e^{ik}]-t_L\tau^{\mathrm R}(k)e^{ik}
-\mu[1+r^{\mathrm R}(k)],\\
(-2t\cos k-\mu)\tau^{\mathrm R}(k)e^{ik}
={}&-t_R[1+r^{\mathrm R}(k)]-t\tau^{\mathrm R}(k)e^{2ik}
-\mu\tau^{\mathrm R}(k)e^{ik}.
\end{aligned}
\end{equation}
Equivalently,
\begin{equation}
t\tau^{\mathrm R}(k)=t_R[1+r^{\mathrm R}(k)],
\qquad
1+r^{\mathrm R}(k)e^{-2ik}=\frac{t_L}{t}\tau^{\mathrm R}(k),
\end{equation}
which yields
\begin{equation}
\begin{aligned}
r^{\mathrm R}(k)&=\frac{t_Rt_L-t^2}{t^2e^{-2ik}-t_Rt_L},
&
\tau^{\mathrm R}(k)&=\frac{tt_R(e^{-2ik}-1)}{t^2e^{-2ik}-t_Rt_L}.
\end{aligned}
\label{eq_reflection_general}
\end{equation}
\subsection{Left scattering state}
We next derive the reflection and transmission amplitudes of the left scattering state and verify the generalized conservation relation defined by the biorthogonal scattering states~\cite{XuPseudoHermiticity2023,RyuBiorthogonal2026}. The left state associated with a right eigenstate of energy $E$ satisfies
\begin{equation*}
|\psi^{\mathrm L}\rangle
=\sum_n\psi^{\mathrm L}(n)\hat c_n^\dagger|0\rangle,
\qquad
\hat H_d^{\dagger}|\psi^{\mathrm L}\rangle
=E^{*}|\psi^{\mathrm L}\rangle.
\end{equation*}
For real $k$, the lead dispersion is real, so $E^{*}=E=-2t\cos k-\mu$. The matrix elements of $\mathcal H^{\dagger}$ across the scattering region are $(\mathcal H^{\dagger})_{1,0}=-t_L^{*}$ and $(\mathcal H^{\dagger})_{0,1}=-t_R^{*}$. Projecting the left single-particle eigenvalue equation onto sites $0$ and $1$ therefore gives
\begin{align}
E^{*}\psi^{\mathrm L}(0)&=-t\psi^{\mathrm L}(-1)-t_R^{*}\psi^{\mathrm L}(1)-\mu\psi^{\mathrm L}(0),
\label{eq_left_reflection_boundary_0}\\
E^{*}\psi^{\mathrm L}(1)&=-t_L^{*}\psi^{\mathrm L}(0)-t\psi^{\mathrm L}(2)-\mu\psi^{\mathrm L}(1).
\label{eq_left_reflection_boundary_1}
\end{align}
For a wave incident from the left, we write
\begin{equation}
\psi^{\mathrm L}(n)=
\begin{cases}
e^{ikn}+r^{\mathrm L}(k)e^{-ikn},&n\leq0,\\
\tau^{\mathrm L}(k)e^{ikn},&n\geq1.
\end{cases}
\label{eq_left_scattering_ansatz}
\end{equation}
where $r^{\mathrm L}(k)$ and $\tau^{\mathrm L}(k)$ are the reflection and transmission amplitudes of the left scattering state, respectively. Adjacent to the scattering region,
\begin{equation}
\begin{aligned}
\psi^{\mathrm L}(-1)&=e^{-ik}+r^{\mathrm L}(k)e^{ik},\quad
\psi^{\mathrm L}(0)=1+r^{\mathrm L}(k),\\
\psi^{\mathrm L}(1)&=\tau^{\mathrm L}(k)e^{ik},\quad
\psi^{\mathrm L}(2)=\tau^{\mathrm L}(k)e^{2ik}.
\end{aligned}
\end{equation}
Substituting these expressions into Eqs.~\eqref{eq_left_reflection_boundary_0} and~\eqref{eq_left_reflection_boundary_1} gives
\begin{equation}
\begin{aligned}
(-2t\cos k-\mu)[1+r^{\mathrm L}(k)]
={}&-t[e^{-ik}+r^{\mathrm L}(k)e^{ik}]-t_R^{*}\tau^{\mathrm L}(k)e^{ik}
-\mu[1+r^{\mathrm L}(k)],\\
(-2t\cos k-\mu)\tau^{\mathrm L}(k)e^{ik}
={}&-t_L^{*}[1+r^{\mathrm L}(k)]-t\tau^{\mathrm L}(k)e^{2ik}
-\mu\tau^{\mathrm L}(k)e^{ik}.
\end{aligned}
\end{equation}
Equivalently,
\begin{equation}
t\tau^{\mathrm L}(k)=t_L^{*}[1+r^{\mathrm L}(k)],
\qquad
1+r^{\mathrm L}(k)e^{-2ik}=\frac{t_R^{*}}{t}\tau^{\mathrm L}(k),
\end{equation}
and hence
\begin{equation}
\begin{aligned}
r^{\mathrm L}(k)&=\frac{t_R^{*}t_L^{*}-t^2}{t^2e^{-2ik}-t_R^{*}t_L^{*}},
&
\tau^{\mathrm L}(k)&=\frac{tt_L^{*}(e^{-2ik}-1)}{t^2e^{-2ik}-t_R^{*}t_L^{*}}.
\end{aligned}
\label{eq_left_scattering_amplitudes}
\end{equation}
\subsection{Biorthogonal conservation}
For real $k$, we define the biorthogonal reflection and transmission coefficients by
\begin{equation}
\mathcal R_{\mathrm{bi}}(k)
=[r^{\mathrm L}(k)]^{*}r^{\mathrm R}(k),
\qquad
\mathcal T_{\mathrm{bi}}(k)
=[\tau^{\mathrm L}(k)]^{*}\tau^{\mathrm R}(k).
\label{eq_biorthogonal_scattering_coefficients}
\end{equation}
Equations~\eqref{eq_reflection_general} and~\eqref{eq_left_scattering_amplitudes} then give
\begin{align}
\mathcal R_{\mathrm{bi}}(k)+\mathcal T_{\mathrm{bi}}(k)
&=\frac{(t_Rt_L-t^2)^2+4t^2t_Rt_L\sin^2k}
{(t^2e^{2ik}-t_Rt_L)(t^2e^{-2ik}-t_Rt_L)}\nonumber\\
&=1.
\label{eq_biorthogonal_conservation}
\end{align}
This is the diagonal, left-incident component of the generalized biorthogonal unitarity condition $S_{\mathrm L}^{\dagger}S_{\mathrm R}=I$~\cite{RyuBiorthogonal2026}.

For a Hermitian scattering region, $t_R=t_L^{*}$ and hence $\mathcal H=\mathcal H^{\dagger}$. The left and right scattering equations are therefore identical. Since the scattering states in Eqs.~\eqref{eq_reflection_ansatz} and~\eqref{eq_left_scattering_ansatz} have the same incident-wave normalization, their reflection and transmission amplitudes coincide,
\begin{equation}
r^{\mathrm L}(k)=r^{\mathrm R}(k)\equiv r(k),
\qquad
\tau^{\mathrm L}(k)=\tau^{\mathrm R}(k)\equiv\tau(k).
\end{equation}
Substituting these relations into Eq.~\eqref{eq_biorthogonal_scattering_coefficients} gives
\begin{align}
\mathcal R_{\mathrm{bi}}(k)
&=[r^{\mathrm L}(k)]^{*}r^{\mathrm R}(k)
=[r(k)]^{*}r(k)=|r(k)|^2\equiv R(k),\nonumber\\
\mathcal T_{\mathrm{bi}}(k)
&=[\tau^{\mathrm L}(k)]^{*}\tau^{\mathrm R}(k)
=[\tau(k)]^{*}\tau(k)=|\tau(k)|^2\equiv T(k).
\end{align}
Thus, the biorthogonal coefficients reduce to the standard reflection and transmission coefficients,
and Eq.~\eqref{eq_biorthogonal_conservation} recovers the conventional probability conservation relation $R(k)+T(k)=1$.

We now restrict $t_R$ and $t_L$ to be real, as considered in the main text. At half filling, the left and right amplitudes reduce to
\begin{equation}
r^{\mathrm L}(k_F)=r^{\mathrm R}(k_F)\equiv r
=\frac{1-t_Rt_L}{1+t_Rt_L},
\qquad
\tau^{\mathrm L}(k_F)=\frac{2t_L}{1+t_Rt_L},
\qquad
\tau^{\mathrm R}(k_F)=\frac{2t_R}{1+t_Rt_L}.
\label{eq_biorthogonal_fermi_amplitudes}
\end{equation}
At the Fermi level, we introduce the shorthand biorthogonal scattering coefficients
\begin{equation}
\mathcal R_{\mathrm{bi}}\equiv\mathcal R_{\mathrm{bi}}(k_F)
=r^{*}r=|r|^2,
\qquad
\mathcal T_{\mathrm{bi}}\equiv\mathcal T_{\mathrm{bi}}(k_F)
=[\tau^{\mathrm L}(k_F)]^{*}\tau^{\mathrm R}(k_F)
=\frac{4t_Rt_L}{(1+t_Rt_L)^2},
\end{equation}
and Eq.~\eqref{eq_biorthogonal_conservation} becomes
\begin{equation}
\mathcal R_{\mathrm{bi}}+\mathcal T_{\mathrm{bi}}=1.
\label{eq_biorthogonal_fermi_conservation}
\end{equation}
Here $\mathcal T_{\mathrm{bi}}$ is an LR bilinear quantity rather than the conventional transmission probability $|\tau^{\mathrm R}(k_F)|^2$. In Q1, $t_Rt_L>0$ gives $0\leq\mathcal R_{\mathrm{bi}}<1$ and $\mathcal T_{\mathrm{bi}}\geq0$. In Q2, $t_Rt_L<0$ instead gives $\mathcal R_{\mathrm{bi}}>1$ and $\mathcal T_{\mathrm{bi}}<0$. Thus, the generalized conservation relation remains valid, while its individual terms no longer admit the positive-probability interpretation of unitary scattering.

Returning to the boundary-induced correlation in Eq.~\eqref{eq_corr_approx}, its algebraically decaying part can be written as
\begin{equation}
C(l+m)\simeq
-r\frac{\sin[k_F(l+m)]}{\pi(l+m)}.
\label{eq_reflection_correlation}
\end{equation}
Therefore, the coefficient of the algebraically decaying Hankel term is exactly $-r$. The minus sign follows from the convention for the boundary image term and does not affect the biorthogonal reflection coefficient $\mathcal R_{\mathrm{bi}}=r^{*}r$. The condition $\mathcal R_{\mathrm{bi}}>1$ thus provides a direct scattering characterization of the nonunitary boundary-critical regime.

\section{The bound states of 1D chain with a general boundary condition}\label{appendixE}
In this part, we analytically obtain the expression of bound states induced by the general boundary condition. First, we propose an ansatz wavefunction $\psi^{R}_{z}(n)=c_{1}z^{n}+c_{2}z^{-n}$ like the discussion before, where $z$ and $z^{-1}$ are two roots of energy spectrum with a certain energy and $|z|<1$. To explicitly show the bound states localized at the position of the general boundary condition, we rewrite the ansatz wavefunction as $\psi^{R}_{z}(n)=a_{1}z^{n}+a_{2}z^{N-n+1}$, where $a_{1}=c_{1}$ and $a_{2}=c_{2}z^{-(N+1)}$. Then, inserting the ansatz wavefunction into Eq.~\eqref{eq_def_cond}, we have the coefficient matrix of $a_{1,2}$:
\begin{equation}
  \mathfrak{A}=
  \begin{pmatrix}
    t-z^{N}t_{R} & z(t z^{N}-t_{R}) \\
    z(t z^{N}-t_{L})& t -z^{N}t_{L}
  \end{pmatrix}.
\end{equation}
Due to $|z|<1$, then we set $z^{N}=0$. So, to ensure $\det(\mathfrak{A})=0$, we find $z=\pm\sqrt{1/t_{R}t_{L}}$, where we set $t=1$. Therefore, we find that if $|t_{R}t_{L}|>1$, the general boundary condition always induces two bound states. Furthermore, solving the coefficient equation, the analytical expression of the right eigenvectors of the bound state is represented as:
\begin{equation}
  \psi_{z}^{R}(n)=a_{1} z^{n} +a_{2}z^{N-n+1},
\end{equation}
where $a_{1}=z t_{R}$ and $a_{2}=1$. Meanwhile, we also use $\psi^{L}_{z}(n)=a_{1}(z^{*})^{n}+a_{2}(z^{*})^{N-n+1}$ to solve Eq.~\eqref{eq_def_cond_left} of the general boundary condition. Then, we obtain a coefficient matrix of the left eigenvectors:
\begin{equation}
  \mathfrak{A}^{L}=
  \begin{pmatrix}
    t-(z^{*})^{N}t^{*}_{L} & z^{*}(t (z^{*})^{N}-t^{*}_{L}) \\
    z^{*}(t (z^{*})^{N}-t^{*}_{R})& t -(z^{*})^{N}t^{*}_{R}
  \end{pmatrix}.
\end{equation}
Similar to the discussion of the right eigenvectors,
the left eigenvector of the bound state is represented as:
\begin{equation}
  \psi_{z}^{L}(n)=a_{1} (z^{*})^{n} +a_{2}(z^{*})^{N-n+1},
\end{equation}
where $a_{1}=z^{*} t^{*}_{L}$ and $a_{2}=1$. Additionally, we also need to bi-normalize the right and left eigenvectors of the bound states, where we obtain the normalization constant $\mathcal{N}^{b}_{RL}=\frac{2 z^{2}}{1-z^{2}}$ through our calculation. Finally, we study the contribution of the bound states for the correlation function $C(l,m)$. Here we only consider $0<z=\sqrt{1/t_{R}t_{L}}<1$, namely $t_{R}t_{L}>1$. Then, the energy $E=-(z^{-1}+z)<\mu$ is occupied, where we take $\mu=0$. Therefore, the bound state contributes to the correlation function $C_{b}(l,m)$, written  as 
\begin{equation}
  C_{b}(l,m)= \frac{(\psi_{z}^{L}(l))^{*}\psi_{z}^{R}(m)}{\mathcal{N}^{b}_{RL}} 
  =z^{l+m}\frac{1-z^{2}}{2z^{2}} = \frac{1}{2}(t_{R}t_{L}-1)(t_{R}t_{L})^{\frac{-(l+m)}{2}}.
\end{equation}

\twocolumngrid

%

\end{document}